\documentclass[aps,twocolumn,preprintnumbers,amsmath,amssymb,nofootinbib,superscriptaddress,notitlepage]{revtex4}
\usepackage{graphicx,color,dcolumn,booktabs,bm}
\usepackage{longtable,lscape}
\usepackage{txfonts}
\usepackage{overpic}
\usepackage{amssymb}
\usepackage{gensymb}
\usepackage{indentfirst}
\usepackage{feynmf}
\usepackage{slashed}
\usepackage{cases}
\usepackage{multirow}
\usepackage{appendix}
\usepackage[figuresright]{rotating}
\usepackage[colorlinks,
citecolor=blue,
anchorcolor=red,
menucolor=red,
linkcolor=red,
filecolor=red,
runcolor=red,
urlcolor=blue,
frenchlinks=red]{hyperref}
\usepackage{epstopdf}
\usepackage{bm}
\usepackage{tabularx}
\usepackage{threeparttable}
\usepackage{bbding}
\usepackage{array}
\usepackage[section]{placeins}
\usepackage{makecell}
\usepackage{comment}
\makeatletter

\newcommand{\Rmnum}[1]{\expandafter\@slowromancap\romannumeral #1@}
\makeatother

\usepackage{soul}
\begin{document}
\title{Role of $4S$-$3D$ mixing in explaining the $\omega$-like $Y(2119)$ observed in\\ $e^+e^-\to\rho\pi$ and $\rho(1450)\pi$}

\author{Zi-Yue Bai}\email{baizy15@lzu.edu.cn}
\affiliation{School of Physical Science and Technology, Lanzhou University, Lanzhou 730000, China}
\affiliation{Lanzhou Center for Theoretical Physics,
Key Laboratory of Theoretical Physics of Gansu Province,
Key Laboratory of Quantum Theory and Applications of MoE,
Gansu Provincial Research Center for Basic Disciplines of Quantum Physics, Lanzhou University, Lanzhou 730000, China}
\affiliation{Research Center for Hadron and CSR Physics, Lanzhou University and Institute of Modern Physics of CAS, Lanzhou 730000, China}

\author{Qin-Song Zhou}\email{zhouqs@imu.edu.cn}
\affiliation{Lanzhou Center for Theoretical Physics,
Key Laboratory of Theoretical Physics of Gansu Province,
Key Laboratory of Quantum Theory and Applications of MoE,
Gansu Provincial Research Center for Basic Disciplines of Quantum Physics, Lanzhou University, Lanzhou 730000, China}
\affiliation{MoE Frontiers Science Center for Rare Isotopes, Lanzhou University, Lanzhou 730000, China}
\affiliation{Center for Quantum Physics and Technologies, School of Physical Science and Technology, Inner Mongolia University, Hohhot 010021, China}

\author{Xiang Liu}\email{xiangliu@lzu.edu.cn}
\affiliation{School of Physical Science and Technology, Lanzhou University, Lanzhou 730000, China}
\affiliation{Lanzhou Center for Theoretical Physics,
Key Laboratory of Theoretical Physics of Gansu Province,
Key Laboratory of Quantum Theory and Applications of MoE,
Gansu Provincial Research Center for Basic Disciplines of Quantum Physics, Lanzhou University, Lanzhou 730000, China}
\affiliation{Research Center for Hadron and CSR Physics, Lanzhou University and Institute of Modern Physics of CAS, Lanzhou 730000, China}
\affiliation{MoE Frontiers Science Center for Rare Isotopes, Lanzhou University, Lanzhou 730000, China}

\date{\today}
\begin{abstract}

The BESIII Collaboration has recently reported a new resonance structure in the cross-section analyses of the $e^+e^- \to \rho\pi$ and $e^+e^- \to \rho(1450)\pi$ processes, displaying characteristics akin to an $\omega$-meson. However, its measured mass, $M = 2119 \pm 11 \pm 15 \, \text{MeV}$, significantly deviates from the predictions of the unquenched relativized potential model for conventional $\omega$ vector meson spectroscopy. To address this discrepancy, we propose a $4S$-$3D$ mixing scheme and investigate the corresponding decay properties within this framework. Our analysis demonstrates that the mixed state $\omega_{4S\text{-}3D}^\prime$ exhibits excellent agreement with the observed resonance, not only in mass and width but also in the products of its dielectron width and branching ratios, $\Gamma_{e^+e^-}^{\mathcal{R}}\times\mathcal{B}_{\mathcal{R}\to\rho\pi}$ and $\Gamma_{e^+e^-}^{\mathcal{R}}\times\mathcal{B}_{\mathcal{R}\to\rho(1450)\pi\to\pi^+\pi^-\pi^0}$. The determined mixing angle, $\theta = -\left(31.1^{+2.1}_{-3.1}\right)\degree$, strongly supports the interpretation of this structure as the $\omega_{4S\text{-}3D}^\prime$ state. Furthermore, we predict dominant decay channels for $\omega_{4S\text{-}3D}^\prime$ and its partner state $\omega_{4S\text{-}3D}^{\prime\prime}$. These predictions, together with the proposed mixing mechanism, provide crucial guidance for future experimental studies aimed at probing this structure and rigorously testing the $4S$-$3D$ mixing hypothesis.

\end{abstract}

\maketitle
\thispagestyle{empty} %

\section{introduction}\label{sec1}
Recently, the BESIII Collaboration measured the Born cross sections for the processes $e^+e^-\to \rho\pi$ and $\rho(1450)\pi$, reporting a resonance structure with a mass of $M=2119\pm 11\pm 15$ MeV and a width of $\Gamma=69\pm 30\pm 5$ MeV \cite{BESIII:2024okl}. This resonance, with a significance of $5.9\sigma$, is hereafter referred to as the $Y(2119)$. The decay channels of the $Y(2119)$ suggest it is a typical $\omega$-like mesonic state with isospin $I=0$. This discovery provides valuable insight into the construction of the $\omega$ meson family. Indeed, this topic has garnered significant interest in recent years \cite{Pang:2019ovr,Wang:2021gle,Wang:2021abg,Zhou:2022wwk,Wang:2022xxi}, particularly due to the rich phenomena observed in light flavor vector mesonic states around 2 GeV \cite{BESIII:2019ebn,BaBar:2019kds,BESIII:2020gnc,BESIII:2020xmw,BESIII:2021bjn,BESIII:2020kpr,BESIII:2021uni,BESIII:2021aet,BESIII:2022yzp,BESIII:2024qjv}.

For $\omega$ meson spectroscopy, observed states such as $\omega(782)$, $\omega(1420)$, and $\omega(1650)$ are documented in the {\it Review of Particle Physics} (RPP) \cite{ParticleDataGroup:2024cfk}. The quantitative mass spectrum of the $\omega$ meson family has been effectively described using potential models, as detailed in Refs. \cite{Wang:2021gle}, which not only accommodates observed states but also predicts higher-lying ones. Comparing the observed $Y(2119)$ with the predicted mass spectrum of the $\omega$ meson family reveals a significant discrepancy, as illustrated in Table \ref{MGI}. This suggests that the $Y(2119)$ cannot be readily classified within the $\omega$ meson family. This intriguing deviation motivates further exploration of the nature of the $Y(2119)$, as observed by BESIII.

Upon examining the details of the above comparison, we observe that the mass of the $\omega(4S)$ is higher than that of the $Y(2119)$. To address this discrepancy, it is essential to identify a mechanism capable of lowering the mass of the $\omega(4S)$ to align with that of the $Y(2119)$. A natural solution is to introduce a $4S$-$3D$ state mixing scheme.
Notably, the $4S$-$3D$ mixing scheme has been successfully applied in studies of the charmonium-like state $Y(4220)$ and the bottomonium-like state $Y(10753)$, as proposed by the Lanzhou group in Refs. \cite{Wang:2019mhs,Wang:2022jxj,Wang:2023zxj,Peng:2024xui,Li:2021jjt,Bai:2022cfz,Li:2022leg,Liu:2024ets,Liu:2023gtx}. Given its effectiveness and potential as a universal mechanism, it is reasonable to extend this approach to the $\omega$ meson family. In the following section, we provide a detailed exploration of the $4S$-$3D$ state mixing scheme as applied to the $\omega$-like state $Y(2119)$.

\section{The $4S$-$3D$ MIXING SCHEME for the $Y(2119)$}
\label{Sec2}

Before introducing the $4S$-$3D$ mixing scheme for the $Y(2119)$, it is crucial to first discuss the mass spectrum of the $\omega$ meson family. Following the approach in Ref. \cite{Wang:2021gle}, one employs a modified Godfrey-Isgur (MGI) potential model, which incorporates relativistic and unquenched effects, to calculate the mass spectrum. By accurately reproducing the masses of observed light-flavor mesons, the model parameters are determined and subsequently used to predict the spectroscopic behavior of high-lying $\omega$ mesonic states. Further details can be found in Ref. \cite{Wang:2021gle}. 

In Table \ref{MGI}, we compile the resonance parameters of several $\omega$ mesonic states around 2 GeV, specifically $\omega(2D)$, $\omega(4S)$, and $\omega(3D)$, and compare them with the experimentally measured mass of the $Y(2119)$. This comparison reveals a discrepancy between the experimental observation and theoretical predictions. Notably, the mass of the $Y(2119)$ is approximately 61 MeV lower than that of the predicted $\omega(4S)$. This inconsistency suggests that categorizing the $Y(2119)$ as a pure $\omega(4S)$ state is problematic. Therefore, a mechanism is required to reconcile the mass difference between the pure $\omega(4S)$ state and the observed $Y(2119)$.

In the charmonium family, $S$-$D$ mixing is a well-established phenomenon, as exemplified by the $\psi(3686)$ and $\psi(3770)$ states, where a $2S$-$1D$ mixing scheme has been identified \cite{Rosner:2001nm}. Inspired by this, a $4S$-$3D$ mixing scheme has been proposed to explain the properties of the charmonium-like state $Y(4220) \equiv \psi(4220)$, with the prediction of its partner state $\psi(4380)$ \cite{Wang:2019mhs}. This mixing scheme has also been successfully applied to decode the bottomonium-like state $Y(10753)$, where a similar $4S$-$3D$ mixing is observed \cite{Li:2021jjt}.

\begin{table}[htbp]
\centering
\caption{The predicted spectroscopic behavior of some high-lying  $\omega$ meson mass spectrum from Ref. \cite{Wang:2021gle} and the comparison with the observed $Y(2119)$ \cite{BESIII:2024okl}.}	\label{MGI}
		\renewcommand\arraystretch{1.5}
		\setlength{\tabcolsep}{4pt}
		\setlength{\arrayrulewidth}{0.5pt}
  \resizebox{\linewidth}{!}{
		\begin{tabular}{c|ccccc}
			\toprule[1pt]
			\toprule[0.5pt]
			$\omega$ states &$\omega(2D)$ &$\omega(4S)$ &$\omega(3D)$ &$Y(2119)$\\
			\midrule[0.5pt]
			Mass (MeV)      &2003         &2180         &2283        &$2119\pm11\pm15$\\
                Width (MeV)     &181          &104          &94          &$69\pm30\pm5$\\
                $\Gamma_{e^+e^-}$ (eV)  &2.2     &7.0          &1.8         &$-$\\
			\bottomrule[0.5pt]
			\bottomrule[1pt]
		\end{tabular}}
	\end{table}	

Motivated by these successes in hadron spectroscopy, we propose a $4S$-$3D$ mixing scheme to address the low-mass problem of the observed $Y(2119)$. This mixing scheme is expressed as:
\begin{align}
\label{mix}
	\left(\begin{array}{c}
	|\omega_{4S-3D}^\prime\rangle \\
	|\omega_{4S-3D}^{\prime\prime}\rangle \end{array}\right)=
	\left(\begin{array}{cc}\cos\,\theta&\sin\,\theta\\
	-\sin\,\theta&\cos\,\theta\end{array}\right)
	\left(\begin{array}{c}|\omega(4S)\rangle\\
	|\omega(3D)\rangle
	\end{array}\right),
\end{align}
where $\theta$ denotes the mixing angle.

The mass eigenvalues of the mixed states are given by
\begin{align}
\label{massmix}
\begin{split}
m_{\omega_{4S-3D}^\prime}^2=&\frac{1}{2}\bigg(m_{\omega(4S)}^2+m_{\omega(3D)}^2\\
&-\sqrt{(m_{\omega(4S)}^2-m_{\omega(3D)}^2)^2\sec^2{2\theta}}\bigg),
\end{split}\\
\begin{split}
m_{\omega_{4S-3D}^{\prime\prime}}^2=&\frac{1}{2}\bigg(m_{\omega(4S)}^2+m_{\omega(3D)}^2\\
&+\sqrt{(m_{\omega(4S)}^2-m_{\omega(3D)}^2)^2\sec^2{2\theta}}\bigg),
\end{split}
\end{align}
where $m_{\omega(4S)}$ and $m_{\omega(3D)}$ are the masses of the pure $\omega(4S)$ and $\omega(3D)$ states, respectively, as listed in Table \ref{MGI}.

\begin{figure}[htbp]
{\centering
\includegraphics[width=0.45\textwidth]{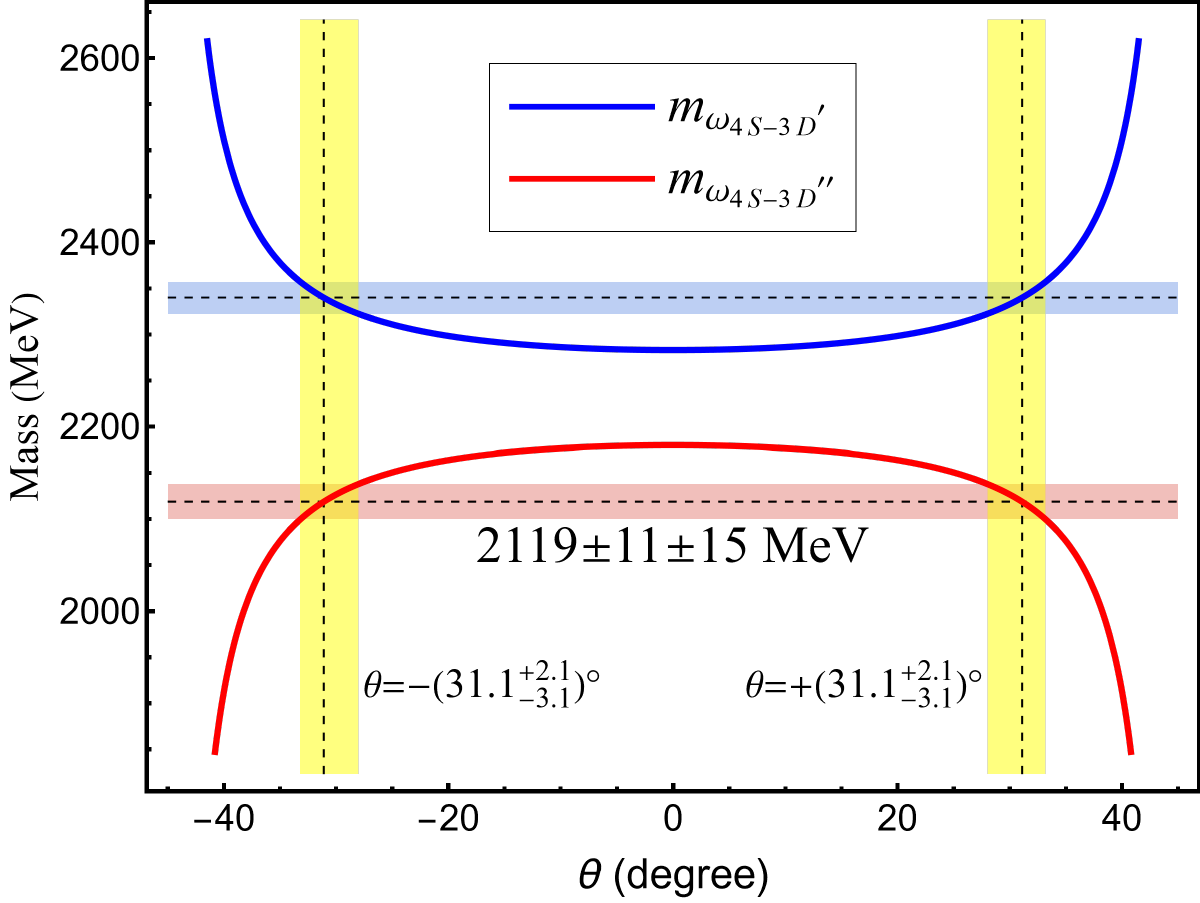}}
\caption{The masses of the mixed state $\omega_{4S-3D}^\prime$ and $\omega_{4S-3D}^{\prime\prime}$ as function of the $4S$-$3D$ mixing angle $\theta$. The light red horizontal band represents the measured mass of the $Y(2119)$, while the yellow vertical bands indicate the mixing angles at which the theoretical mass of $\omega_{4S-3D}^\prime$ coincides with the $Y(2119)$. Additionally, the light blue horizontal band represents the predicted mass of $\omega_{4S-3D}^{\prime\prime}$ at these mixing angles.}
\label{figmix}
\end{figure}

Using the measured mass of the $Y(2119)$, $M=2119\,\pm11\,\pm15$ MeV, as input for $m_{\omega_{4S-3D}^\prime}$, we determine the mixing angle to be 
\begin{align}
\theta=\pm(31.1^{+2.1}_{-3.1})\degree,
\end{align}
as illustrated in Fig. \ref{figmix}.
Next, we analyze the total width, the partial widths for the $\rho\pi$ and $\rho(1450)\pi$ channels, as well as the dielectron width of the mixed ${\omega_{4S-3D}^\prime}$ states for $\theta=\pm\left(31.1^{+2.1}_{-3.1}\right)\degree$.

The strong decays of $\omega_{4S-3D}^\prime$ and $\omega_{4S-3D}^{\prime\prime}$, allowed by the Okubo-Zweig-Iizuka (OZI) rule, are calculated using the quark pair creation (QPC) model \cite{Micu:1968mk,LeYaouanc:1977gm,Jacob:1959at}, with model parameters identical to those in Refs. \cite{Wang:2021abg,Wang:2021gle}. The methodology follows the approach outlined in Refs. \cite{Wang:2021abg,Wang:2021gle}, but differs in the amplitude formulation, which incorporates contributions from both $4S$-wave and $3D$-wave components. The decay widths of $\omega_{4S-3D}^{\prime}$ and $\omega_{4S-3D}^{\prime\prime}$ in the rest frame of the initial state are given by
\begin{align}
\begin{split}
\Gamma_{\omega_{4S-3D}^\prime \to BC} = & \frac{\pi}{4} \frac{|\mathbf{P}|}{m_{\omega_{4S-3D}^\prime}^2} \sum_{JL} \left| \cos\theta \mathcal{M}_{\omega(4S)}^{JL}(\mathbf{P}) \right.  \\
& \quad \left. + \sin\theta \mathcal{M}_{\omega(3D)}^{JL}(\mathbf{P}) \right|^2,
\end{split}
\end{align}
\begin{align}
\begin{split}
\Gamma_{\omega_{4S-3D}^{\prime\prime} \to BC} = & \frac{\pi}{4} \frac{|\mathbf{P}|}{m_{\omega_{4S-3D}^{\prime\prime}}^2} \sum_{JL} \left| -\sin\theta \mathcal{M}_{\omega(4S)}^{JL}(\mathbf{P}) \right.  \\
& \quad \left. + \cos\theta \mathcal{M}_{\omega(3D)}^{JL}(\mathbf{P}) \right|^2,
\end{split}
\end{align}
where $\mathbf{P}=\mathbf{P}_B=-\mathbf{P}_C$, $\textbf{\textit{L}}$ and $\textbf{\textit{J}}$ represent the relative orbital angular and total spin momentum between final states $B$ and $C$. The terms $\mathcal{M}_{\omega(4S)}^{JL}(\mathbf{P})$ and $\mathcal{M}_{\omega(3D)}^{JL}(\mathbf{P})$ denote the partial wave amplitudes of $\omega(4S)\to BC$ and $\omega(3D)\to BC$, respectively.

The dielectron widths of the mixed $\omega_{4S-3D}^\prime$ and $\omega_{4S-3D}^{\prime\prime}$ states can be formulated as described in Refs. \cite{Godfrey:1985xj,Wang:2021abg}

\begin{align}
\Gamma_{e^+e^-}^{\omega_{4S-3D}^\prime} &= \frac{4\pi\alpha^2 m_{\omega_{4S-3D}^\prime}}{3}  \left| \cos\theta \mathcal{M}_{\omega(4S)} + \sin\theta \mathcal{M}_{\omega(3D)} \right|^2, \\
\Gamma_{e^+e^-}^{\omega_{4S-3D}^{\prime\prime}} &= \frac{4\pi\alpha^2 m_{\omega_{4S-3D}^{\prime\prime}}}{3}  \left| -\sin\theta \mathcal{M}_{\omega(4S)}^\prime + \cos\theta \mathcal{M}_{\omega(3D)}^\prime \right|^2,
\label{3Dee}
\end{align}
where 
\begin{align}
\begin{split}
\mathcal{M}_{\omega_{4S-3D}^\prime}=\sqrt{\frac{2}{3}}{\mathcal{A}}_{\omega(4S)},\,\mathcal{M}_{\omega(3D)}=\sqrt{\frac{4}{27}}{\mathcal{A}}_{\omega(3D)},\\
\mathcal{M}_{\omega_{4S-3D}^{\prime\prime}}^\prime=\sqrt{\frac{2}{3}}{\mathcal{A}}_{\omega(4S)}^\prime,\,\mathcal{M}_{\omega(3D)}^\prime=\sqrt{\frac{4}{27}}{\mathcal{A}}_{\omega(3D)}^\prime,
\end{split}
\end{align}
and ${\mathcal{A}}_{\omega(4S)}^{(\prime)}$, ${\mathcal{A}}_{\omega(3D)}^{(\prime)}$ can be written as

\begin{align}
\begin{split}
{\mathcal{A}}_{\omega(4S)}=&(2\pi)^{-3/2}\tilde{m}^{1/2}_{\omega(4S)}m_{\omega_{4S-3D}^\prime}^{-2}\\
&\times\int d^3\mathbf{p}(4\pi)^{-1/2}\Phi(p)\left(\frac{m_1m_2}{E_1E_2}\right)^{1/2},
\end{split}\\
\begin{split}
{\mathcal{A}}_{\omega(3D)}=&(2\pi)^{-3/2}\tilde{m}^{1/2}_{\omega(3D)}m_{\omega_{4S-3D}^\prime}^{-2}\int d^3\mathbf{p}(4\pi)^{-1/2}\\
&\times\phi(p)\left(\frac{m_1m_2}{E_1E_2}\right)^{1/2}\left(\frac{p}{E_1}\right)^2,
\end{split}\\
\begin{split}
{\mathcal{A}}_{\omega(4S)}^\prime=&(2\pi)^{-3/2}\tilde{m}^{1/2}_{\omega(4S)}m_{\omega_{4S-3D}^{\prime\prime}}^{-2}\\
&\times\int d^3\mathbf{p}(4\pi)^{-1/2}\Phi(p)\left(\frac{m_1m_2}{E_1E_2}\right)^{1/2},
\end{split}\\
\begin{split}
{\mathcal{A}}_{\omega(3D)}^\prime=&(2\pi)^{-3/2}\tilde{m}^{1/2}_{\omega(3D)}m_{\omega_{4S-3D}^{\prime\prime}}^{-2}\int d^3\mathbf{p}(4\pi)^{-1/2}\\
&\times\phi(p)\left(\frac{m_1m_2}{E_1E_2}\right)^{1/2}\left(\frac{p}{E_1}\right)^2,
\end{split}
\end{align}
in which $\alpha\approx1/137$ is the fine-structure constant. $\Phi(p)$ and $\phi(p)$ are the radial spatial wave function for the pure $\omega(4S)$ and $\omega(3D)$ states in the momentum space, respectively. These wave functions can be obtained by solving the unquenched MGI model \cite{Wang:2021abg,Wang:2021gle}, and the complete spatial wave functions are given by $\Phi(\mathbf{p})=\Phi(p)Y_{00}(\mathbf{\Omega_p})$, $\phi(\mathbf{p})=\phi(p)Y_{2m}(\mathbf{\Omega_p})$, where $Y_{lm}(\mathbf{\Omega_p})$ are the spherical harmonics. Here, $m_1,\,m_2$ represent the consistent quark masses of the $\omega$ meson, for which we take $m_1=m_2=220\,\text{MeV}$, The energies of the quarks are given by $E_1=\sqrt{m_1^2+p^2}$, $E_2=\sqrt{m_2^2+p^2}$, with the total energy being $E=E_1+E_2$. $\tilde{m}_{\omega(4S)}$ and $\tilde{m}_{\omega(3D)}$ take the following the form,
\begin{align}
\begin{split}
\tilde{m}_{\omega(4S)}=2\int d^3\mathbf{p}E|\Phi(\mathbf{p})|^2,\\
\tilde{m}_{\omega(3D)}=2\int d^3\mathbf{p}E|\phi(\mathbf{p})|^2.
\end{split}
\end{align}

Using the formulas introduced above, we obtain the resonance parameters, partial widths for the $\rho\pi$ and $\rho(1450)\pi$ channels, as well as the dielectron width of the mixed $\omega_{4S-3D}^\prime$ for $\theta=\pm(31.1^{+2.1}_{-3.1})\degree$. These values are summarized in Table \ref{comparison}. With the measured mass of the $Y(2119)$ as input, we obtain a total width of $66.9^{+11.4}_{-9.9}$ MeV for the negative mixing angle, and $52.9^{+11.5}_{-9.5}$ MeV for the positive mixing angle. These results are consistent with the measured width of the $Y(2119)$, $\Gamma = 69 \pm 30 \pm 5$ MeV \cite{BESIII:2024okl}. This agreement supports the identification of the $Y(2119)$ as the mixed $\omega_{4S-3D}^\prime$ state.

\begin{table*}[htbp]
\centering
\caption{Theoretical predictions for the mixed $\omega_{4S\text{-}3D}^\prime$ state's mass, width ($\Gamma$), dielectron width ($\Gamma^{\mathcal{R}}_{e^+e^-}$), partial widths of $\rho\pi$ and $\rho(1450)\pi$ channels, and products $\Gamma_{e^+e^-}^\mathcal{R} \times \mathcal{B}_{\mathcal{R}\to\rho\pi}$ and $\Gamma_{e^+e^-}^\mathcal{R}\times \mathcal{B}_{\mathcal{R}\to\rho(1450)\pi\to\pi^+\pi^-\pi^0}$, calculated for both negative and positive angles. Comparison with BESIII Collaboration measurements of the observed $Y(2119)$ for constructive and destructive interference scenarios \cite{BESIII:2024okl} are included.} 
\label{comparison}
\renewcommand\arraystretch{1.5}
\setlength{\tabcolsep}{9pt}
\setlength{\arrayrulewidth}{0.5pt}
\resizebox{\linewidth}{!}{
\begin{tabular}{c|c|c|c|c}
\toprule[1pt]
\toprule[0.5pt]
Intermediate state $\mathcal{R}$ &\multicolumn{2}{c|}{$\omega_{4S-3D}^\prime$}  &\multicolumn{2}{c}{$Y(2119)$  \cite{BESIII:2024okl}} \\
\midrule[0.5pt]
Status  &$\theta$=$-(31.1^{+2.1}_{-3.1})\degree$ &$\theta$=$+(31.1^{+2.1}_{-3.1})\degree$ &Constructive interference scenario &Destructive interference scenario \\
\midrule[0.5pt]
Mass (MeV) &$2119\pm11\pm15$  &$2119\pm11\pm15$   &$2119\pm11\pm15$  &$2119\pm11\pm15$ \\
$\Gamma$ (MeV)     &$66.9^{+11.4}_{-9.9}$ &$52.9^{+11.5}_{-9.5}$  &$69\pm30\pm5$ &$69\pm30\pm5$\\
$\Gamma_{e^+e^-}^{\mathcal{R}}$ (eV) &$10.0\pm0.2$ &$2.6^{+0.4}_{-0.2}$  &$-$  &$-$\\
$\Gamma_{\mathcal{R}\to\rho\pi}$ (MeV) &$12.8^{+2.2}_{-1.9}$ &$4.6^{+1.4}_{-1.0}$ &$-$  &$-$\\
$\Gamma_{\mathcal{R}\to\rho(1450)\pi}$ (MeV) &$24.8^{+4.4}_{-4.1}$ &$8.5^{+3.0}_{-2.0}$ &$-$  &$-$\\
$\Gamma_{e^+e^-}^{\mathcal{R}}\times {\mathcal{B}}_{\mathcal{R}\to\rho\pi}\,(\text{eV})$ &$1.9\pm0.5$ &$0.2\pm0.1$ &$1.5\pm0.7$ &$17\pm12$\\
$\Gamma_{e^+e^-}^{\mathcal{R}}\times {\mathcal{B}}_{\mathcal{R}\to\rho(1450)\pi\to\pi^+\pi^-\pi^0}\,(\text{eV})$ &$0.4\pm0.1$ &$0.04\pm0.02$ &$0.3\pm0.2$ &$3\pm2$ \\
\bottomrule[0.5pt]
\bottomrule[1pt]
\end{tabular}}
\end{table*}

Except for the resonance parameters, the BESIII Collaboration also extracted the products of the dielectron width and branching ratios for the $Y(2119)$ in the processes $e^+e^- \to \rho\pi$ and $e^+e^- \to \rho(1450)\pi$, denoted as $\Gamma_{e^+e^-}^{Y(2119)}\times {\mathcal{B}}_{Y(2119)\to\rho\pi}$ and $\Gamma_{e^+e^-}^{Y(2119)}\times {\mathcal{B}}_{Y(2119)\to\rho(1450)\pi\to\pi^+\pi^-\pi^0}$, respectively. These quantities were obtained by fitting the measured cross section data. Two solutions were found, sharing the same resonance parameters but differing in the relative phase, which indicates constructive and destructive interference between the resonance amplitudes and the continuum background.

To further examine the hypothesis that the $Y(2119)$ corresponds to the mixed $\omega_{4S-3D}^\prime$ state, a comparison between the experimental measurements and our theoretical estimations of the products of the dielectron width and branching ratios for the $\rho\pi$ and $\rho(1450)\pi$ processes is essential.

By utilizing the decay information obtained from our calculations, as detailed in Table \ref{comparison}, we are able to derive the physical quantity
$\Gamma_{e^+e^-}^{\omega_{4S-3D}^\prime}\times {\mathcal{B}}_{\omega_{4S-3D}^\prime\to\rho\pi}$.
To determine the magnitude of $\Gamma_{e^+e^-}^{\omega_{4S-3D}^\prime}\times {\mathcal{B}}_{\omega_{4S-3D}^\prime\to\rho(1450)\pi\to\pi^+\pi^-\pi^0}$, we first need to establish theoretical inputs for the $\rho(1450)$ state. Given the significant variation in the resonance parameters for $\rho(1450)$ as listed by the Particle Data Group (PDG) \cite{ParticleDataGroup:2024cfk}, we assume the $\rho(1450)$ to be the $\rho(2S)$ state and proceed to determine its mass and decay properties within the theoretical frameworks of the MGI and QPC models.

We then compare these theoretical results with available experimental measurements, which are summarized in Table \ref{rho(1450)}. Our calculated resonance parameters, as well as several related ratios for $\Gamma_{\rho(1450)\to\pi\pi}$, show good consistency with some experimental values. Based on this comparison, we identify the $\rho(1450)$ as the $\rho(2S)$ state and proceed with further calculations. The theoretical branching ratio of $\rho(1450)\to\pi\pi$ that we adopt is
\begin{align}
    {\mathcal{B}}(\rho(1450)\to\pi\pi)=6.7/64.4=10.4\%.
\end{align}

\begin{table}[htbp]
\centering
\caption{Theoretical predictions for the mass, width ($\Gamma$), dielectron width ($\Gamma_{e^+e^-}$), partial width of $\pi\pi$ ($\Gamma_{\pi\pi}$), and several related ratios of $\Gamma_{\rho(1450)\to\pi\pi}$ for the $\rho(1450)$ identified as $\rho(2S)$. Corresponding experimental measurements are also provided for comparison.}

\label{rho(1450)}
\renewcommand\arraystretch{1.5}
\setlength{\tabcolsep}{11pt}
\setlength{\arrayrulewidth}{0.5pt}
\resizebox{\linewidth}{!}{
\begin{tabular}{c|cc}
\toprule[1pt]
\toprule[0.5pt]
$\rho(1450)$   &Theory  &Experiment \\
\midrule[0.5pt]
\multirow{7}{*}{Mass (MeV)} &\multirow{7}{*}{1413}  &$1421\pm15$ \cite{CMD-2:2000mlo} \\
&{} &$1435\pm40$ \cite{CRYSTALBARREL:2001ldq}\\
&{} &$1429\pm41$ \cite{BaBar:2017dwm}\\
&{} &$1406\pm15$ \cite{CLEO:1999dln}\\
&{} &$1411\pm14$ \cite{CrystalBarrel:1997mmp}\\
&{} &$1424\pm25$ \cite{DM2:1988xqd}\\
&{} &$1422.8\pm6.5$ \cite{CrystalBarrel:1999sjf}\\
$\Gamma$ (MeV)     &64.4 &$60\pm15$ \cite{Fukui:1988mp} \\
$\Gamma_{e^+e^-}$ (keV) &0.387 &$-$  \\
$\Gamma_{\pi\pi}$ (MeV) &6.7   &$-$ \\
$\Gamma_{\pi\pi}/\Gamma_{\omega\pi}$ &0.15  &$\sim0.32$ \cite{Clegg:1993mt} \\
$\Gamma_{\pi\pi}/\Gamma_{\eta\rho}$ &1.1  &$1.3\pm0.4$ \cite{SND:2014rfi} \\
$\Gamma_{\pi\pi}\times\Gamma_{e^+e^-}/\Gamma$ (keV) &0.040  &$0.027^{+0.015}_{-0.010}$ \cite{Kurdadze:1983ys}\\
\bottomrule[0.5pt]
\bottomrule[1pt]
\end{tabular}}
\end{table}

Finally, we derive the products of the dielectron width and branching ratios, $\Gamma_{e^+e^-}^{\omega_{4S-3D}^\prime} \times \mathcal{B}_{\omega_{4S-3D}^\prime \to \rho\pi}$ and $\Gamma_{e^+e^-}^{\omega_{4S-3D}^\prime} \times \mathcal{B}_{\omega_{4S-3D}^\prime \to \rho(1450)\pi \to \pi^+\pi^-\pi^0}$, from theoretical calculations. These results are compared with the corresponding values for the experimentally measured $Y(2119)$, as presented in Table~\ref{comparison}. 

As shown in Table~\ref{comparison}, the mixed state $\omega_{4S-3D}^\prime$ with a mixing angle $\theta = \pm(31.1^{+2.1}_{-3.1})^\circ$ is consistent with the resonance parameters of the $Y(2119)$ obtained from experimental measurements. The products of the dielectron width and branching ratios for the observed channels provide further distinction between the two cases. Specifically, the $\omega_{4S-3D}^\prime$ state with a negative mixing angle yields products of the dielectron width and branching ratios, $\mathcal{B}_{\mathcal{R} \to \rho\pi}$ and $\mathcal{B}_{\mathcal{R} \to \rho(1450)\pi \to \pi^+\pi^-\pi^0}$, that are in excellent agreement with those of the measured $Y(2119)$ in the constructive interference scenario. This suggests that our assignment of the $Y(2119)$ state to the mixed state $\omega_{4S-3D}^\prime$ with $\theta = -(31.1^{+2.1}_{-3.1})^\circ$ is reasonable. Furthermore, we argue that the constructive interference between the resonance amplitude and the continuum background for the $Y(2119)$ is physically valid, whereas the destructive case does not correspond to a viable physical solution.

Based on the preceding analyses, we conclude that our proposed $4S$-$3D$ mixing scheme for the $Y(2119)$ structure in the processes $e^+e^- \to \rho\pi$ and $\rho(1450)\pi$ is well-supported, with a determined mixing angle of $\theta = -(31.1^{+2.1}_{-3.1})^\circ$. In addition to the $\omega_{4S-3D}^\prime$ state discussed in detail, its partner state $\omega_{4S-3D}^{\prime\prime}$ is predicted via Eq.~\eqref{mix}. Table~\ref{channels} summarizes the resonance parameters and dominant decay channels for both states.

\begin{table}[htbp]
		\centering
		\caption{Spectroscopic properties of two mixed states with $\theta$=$-(31.1^{+2.1}_{-3.1})\degree$.}
		\label{channels}
		\renewcommand\arraystretch{1.3}
		\setlength{\tabcolsep}{9pt}
		\setlength{\arrayrulewidth}{0.5pt}
  \resizebox{\linewidth}{!}{
		\begin{tabular}{c|cc}
			\toprule[1pt]
			\toprule[0.5pt]
			States &$\omega_{4S-3D}^\prime$ &$\omega_{4S-3D}^{\prime\prime}$\\
			\midrule[0.5pt]
			Mass (MeV)      &$2119\pm11\pm15$         &$2340^{+17}_{-18}$ \\
                $\Gamma$ (MeV)     &$66.9^{+11.4}_{-9.9}$  &$147.3^{+14.9}_{-16.0}$   \\
                $\Gamma_{e^+e^-}$ (eV)  &$10.0\pm0.2$          &$0.02^{+0.04}_{-0.02}$   \\
                $\Gamma_{\rho\pi}$ (MeV)  &$12.8^{+2.2}_{-1.9}$ &$1.0^{+0.7}_{-0.6}$\\
                $\Gamma_{\rho(1450)\pi}$ (MeV)  &$24.8^{+4.4}_{-4.1}$ &$1.6\pm1.0$ \\
                $\Gamma_{\rho(1700)\pi}$ (MeV)  &$0.8\pm0.1$ &$0.4\pm0.1$ \\
                $\Gamma_{\rho_3(1690)\pi}$ (MeV)  &$1.9^{+0.5}_{-0.4}$ &0 \\
                $\Gamma_{\rho\pi(1300)}$ (MeV)  &$0.7\pm0.2$ &$1.0^{+0.7}_{-0.6}$\\
                $\Gamma_{\rho a_0(980)}$ (MeV)  &$10.6^{+0.1}_{-0.2}$ &$23.8^{+1.0}_{-0.9}$ \\
                $\Gamma_{\rho a_0(1450)}$ (MeV)  &$-$ &$29.3^{+2.0}_{-3.7}$ \\
                $\Gamma_{\rho a_1(1260)}$ (MeV)  &$4.2^{+1.8}_{-1.4}$ &$14.5\pm1.4$ \\
                $\Gamma_{\rho a_2(1320)}$ (MeV)  &$0.3^{+0.6}_{-0.3}$ &$6.5^{+1.6}_{-0.9}$ \\
                $\Gamma_{b_1(1235)\pi}$ (MeV)  &$7.2^{+0.8}_{-0.7}$ &$50.5^{+3.2}_{-3.6}$ \\
                $\Gamma_{\eta h_1(1170)}$ (MeV)  &$0.7\pm0.1$ &$5.6^{+0.6}_{-0.7}$ \\
                $\Gamma_{\omega f_1(1285)}$ (MeV)  &$0.2^{+0.3}_{-0.1}$ &$3.1\pm0.2$ \\
                $\Gamma_{\omega f_1(1420)}$ (MeV)  &$-$ &$1.3^{+0.2}_{-0.3}$ \\
                $\Gamma_{\omega f_2(1270)}$ (MeV)  &$0.4^{+0.3}_{-0.2}$ &$3.5^{+1.0}_{-0.7}$ \\
                $\Gamma_{K\bar K(1460)+h.c.}$ (MeV) &$0$ &$1.6\pm0.2$ \\
                $\Gamma_{K\bar K_1(1270)+h.c.}$ (MeV) &$0.8\pm0.1$ &$0.1\pm0.1$ \\
			\bottomrule[0.5pt]
			\bottomrule[1pt]
		\end{tabular}}
	\end{table}

For the $\omega_{4S-3D}^\prime$ state, significant decay modes include $\rho\pi$, $\rho(1450)\pi$, $\rho a_0(980)$ and $b_1(1235)\pi$. We predict the following products for experimental searches,

\begin{align}
\begin{split}
\Gamma_{e^+e^-}^{\omega_{4S-3D}^\prime}\times {\mathcal{B}}_{\omega_{4S-3D}^\prime\to\rho a_0(980)}=\left(1.6\pm0.3\right)\,\text{eV},\\
\Gamma_{e^+e^-}^{\omega_{4S-3D}^\prime}\times {\mathcal{B}}_{\omega_{4S-3D}^\prime\to b_1(1235)\pi}=\left(1.1^{+0.3}_{-0.2}\right)\,\text{eV}.
\end{split}
\end{align}
These channels offer promising avenues for future experimental test.

In contrast, the $\omega_{4S-3D}^{\prime\prime}$ state exhibits a suppressed dielectron width due to a strong destructive interference between $4S$- and $3D$-wave amplitudes in Eq. (\ref{3Dee}) at $\theta \approx -31\degree$. Its dominant decays are $b_1(1235)\pi$, $\rho a_0(1450)$ and $\rho a_0(980)$, with $b_1(1235)\pi$ being the most viable detection channel. However, the predicted product  
\begin{align}
\Gamma_{e^+e^-}^{\omega_{4S-3D}^{\prime\prime}}\times {\mathcal{B}}_{\omega_{4S-3D}^{\prime\prime}\to b_1(1235)\pi}=\left(0.01^{+0.02}_{-0.01}\right)\,\text{eV}
\end{align}
implies significant experimental challenges in observing this state via $e^+e^-$ {collision}.

\section{SUMMARY}
\label{SecIII}
Investigating the spectroscopic properties of light flavor mesons offers profound insights into the non-perturbative behavior of the strong interaction. The ongoing accumulation of experimental data in the $2.0$--$3.1$ GeV range is crucial for advancing light flavor meson spectroscopy, particularly for highly excited states above $2.0$ GeV \cite{BESIII:2024okl,BESIII:2019ebn,BaBar:2019kds,BESIII:2020gnc,BESIII:2020xmw,BESIII:2021bjn,BESIII:2020kpr,BESIII:2021uni,BESIII:2021aet,BESIII:2022yzp,BESIII:2024qjv}. In this context, not only are the resonance parameters significant, but the line shape and the magnitude of the measured cross section also play a pivotal role in refining theoretical frameworks and deepening our understanding of light flavor meson spectroscopy.

Recently, the BESIII Collaboration observed a new structure, $Y(2119)$, through the study of the cross sections of the $e^+e^- \to \rho\pi$ and $e^+e^- \to \rho(1450)\pi$ processes \cite{BESIII:2024okl}. The decay modes of the $Y(2119)$ suggest that it is an $\omega$-like mesonic state with isospin $I=0$. However, its measured mass deviates significantly from the $\omega$ vector meson spectrum predicted by the unquenched relativized potential model \cite{Wang:2021gle}, being approximately $61$ MeV smaller than the theoretical mass of $\omega(4S)$. Inspired by our previous work on the $Y(4220)$ \cite{Wang:2019mhs} and $\Upsilon(10753)$ \cite{Li:2021jjt}, we propose a $4S$-$3D$ mixing scheme to address this discrepancy. Notably, $S$-$D$ mixing is a universal mechanism observed in both heavy and light flavor meson systems.

Our calculations reveal that when the mixing angle approaches $\pm(31.1^{+2.1}_{-3.1})\degree$, the mass of the mixed state $\omega_{4S\text{-}3D}^\prime$ aligns well with the measured mass of the $Y(2119)$. We further investigate the decay properties for mixing angles $\theta = -(31.1^{+2.1}_{-3.1})\degree$ and $\theta = (31.1^{+2.1}_{-3.1})\degree$, respectively. The results demonstrate that both scenarios yield widths consistent with the measured width of the $Y(2119)$. 

Additionally, we examine the products of the dielectron width and branching ratios, $\Gamma_{e^+e^-}^{\mathcal{R}} \times \mathcal{B}_{\mathcal{R} \to \rho\pi}$ and $\Gamma_{e^+e^-}^{\mathcal{R}} \times \mathcal{B}_{\mathcal{R} \to \rho(1450)\pi \to \pi^+\pi^-\pi^0}$, comparing theoretical predictions with experimental measurements. The comparison indicates that the results for the negative mixing angle agree with the constructive interference scenario of the $Y(2119)$ observed in experiments. This supports our $4S$-$3D$ mixing explanation for the  $Y(2119)$, with a mixing angle of $\theta = -(31.1^{+2.1}_{-3.1})\degree$. It also suggests that the constructive interference scenario of the $Y(2119)$ is physical, while the destructive interference scenario is not.

Furthermore, we predict the dominant decay channels for the $\omega_{4S\text{-}3D}^\prime$ state and its partner state $\omega_{4S\text{-}3D}^{\prime\prime}$. Our results indicate that, in addition to the $\rho\pi$ and $\rho(1450)\pi$ channels, the $\rho a_0(980)$ and $b_1(1235)\pi$ channels are also significant for $\omega_{4S\text{-}3D}^\prime$ and could potentially be observed in experiments such as those conducted by the BESIII Collaboration. For the $\omega_{4S\text{-}3D}^{\prime\prime}$ state, we find that it may be challenging to identify in $e^+e^-$ collision experiments due to its small dielectron width.

These analyses provide valuable guidance for future experiments aimed at studying the $Y(2119)$ state and rigorously testing our $4S$-$3D$ mixing hypothesis.

\vfil
\section*{Acknowledgments}

This work is supported by the National Natural Science Foundation of China under Grant Nos. 12335001, and 12247101,  the ‘111 Center’ under Grant No. B20063, the Natural Science Foundation of Gansu Province (No. 22JR5RA389), the fundamental Research Funds for the Central Universities, and the project for top-notch innovative talents of Gansu province. Q.S. Zhou also acknowledges support from The Research Support Program for High-Level Talents at the Autonomous Region level in the Inner Mongolia Autonomous Region (Grant No. 13100-15112049), and The Research Startup Project of Inner Mongolia University (Grant No. 10000-23112101/101).


\begin{thebibliography}{99}

\bibitem{BESIII:2024okl}
M.~Ablikim \textit{et al.} [BESIII Collaboration],
Study of $e^+e^-\to\pi^+\pi^-\pi^0$ at $\sqrt s$ from 2.00 to 3.08~GeV at BESIII,
\href{https://doi.org/10.1103/PhysRevD.110.032005}{Phys. Rev. D \textbf{110}, 032005 (2024)}.

\bibitem{Pang:2019ovr}
C.~Q.~Pang, Y.~R.~Wang, J.~F.~Hu, T.~J.~Zhang, and X.~Liu,
Study of the $\omega$ meson family and newly observed $\omega$-like state $X(2240)$,
\href{https://doi.org/10.1103/PhysRevD.101.074022}{Phys. Rev. D \textbf{101}, 074022 (2020)}.

\bibitem{Wang:2021gle}
J.~Z.~Wang, L.~M.~Wang, X.~Liu, and T.~Matsuki,
Deciphering the light vector meson contribution to the cross sections of $e^+e^-$ annihilations into the open-strange channels through a combined analysis,
\href{https://doi.org/10.1103/PhysRevD.104.054045}{Phys. Rev. D \textbf{104}, 054045 (2021)}.

\bibitem{Wang:2021abg}
L.~M.~Wang, S.~Q.~Luo, and X.~Liu,
Light unflavored vector meson spectroscopy around the mass range of $2.4\sim3.0$ GeV and possible experimental evidence,
\href{https://doi.org/10.1103/PhysRevD.105.034011}{Phys. Rev. D \textbf{105}, 034011 (2022)}.

\bibitem{Zhou:2022wwk}
Q.~S.~Zhou, J.~Z.~Wang, and X.~Liu,
Role of the $\omega(4S)$ and $\omega(3D)$ states in mediating the $e^+e^-\to\omega\eta$ and $\omega\pi^0\pi^0$ processes,
\href{https://doi.org/10.1103/PhysRevD.106.034010}{Phys. Rev. D \textbf{106}, 034010 (2022)}.

\bibitem{Wang:2022xxi}
Y.~R.~Wang, T.~Y.~Li, Z.~Y.~Fang, H.~Chen, and C.~Q.~Pang,
Study of the \ensuremath{\omega} and $\omega_3$, \ensuremath{\rho} and $\rho_3$, and the newly observed \ensuremath{\omega}-like state X(2220),
\href{https://doi.org/10.1103/PhysRevD.106.114024}{Phys. Rev. D \textbf{106}, 114024 (2022)}.

\bibitem{BESIII:2019ebn}
M.~Ablikim \textit{et al.} [BESIII Collaboration],
Cross section measurements of $e^{+}e^{-} \to K^{+}K^{-}K^{+}K^{-} $ and $ \phi K^{+}K^{-}$ at center-of-mass energies from 2.10 to 3.08 GeV,
\href{https://doi.org/10.1103/PhysRevD.100.032009}{Phys. Rev. D \textbf{100}, 032009 (2019)}.

\bibitem{BaBar:2019kds}
J.~P.~Lees \textit{et al.} [BaBar Collaboration],
Resonances in $e^+e^-$ annihilation near 2.2 GeV,
\href{https://doi.org/10.1103/PhysRevD.101.012011}{Phys. Rev. D \textbf{101}, 012011 (2020)}.

\bibitem{BESIII:2020gnc}
M.~Ablikim \textit{et al.} [BESIII Collaboration],
Observation of a structure in $e^{+}e^{-} \to \phi \eta^{\prime}$ at $\sqrt{s}$ from 2.05 to 3.08 GeV,
\href{https://doi.org/10.1103/PhysRevD.102.012008}{Phys. Rev. D \textbf{102}, 012008 (2020)}.

\bibitem{BESIII:2020xmw}
M.~Ablikim \textit{et al.} [BESIII Collaboration],
Observation of a resonant structure in $e^{+}e^{-} \to \omega\eta$ and another in $e^{+}e^{-} \to \omega\pi^{0}$ at center-of-mass energies between 2.00 and 3.08 GeV,
\href{https://doi.org/10.1016/j.physletb.2020.136059}{Phys. Lett. B \textbf{813}, 136059 (2021)}.

\bibitem{BESIII:2021bjn}
M.~Ablikim \textit{et al.} [BESIII Collaboration],
Study of the process $e^{+}e^{-}\rightarrow\phi\eta$ at center-of-mass energies between 2.00 and 3.08 GeV,
\href{https://doi.org/10.1103/PhysRevD.104.032007}{Phys. Rev. D \textbf{104}, 032007 (2021)}.

\bibitem{BESIII:2020kpr}
M.~Ablikim \textit{et al.} [BESIII Collaboration],
Measurement of the Born cross sections for $e^+e^- \to \eta^\prime \pi^{+}\pi^{-}$ at center-of-mass energies between $2.00$ and $3.08$ GeV,
\href{https://doi.org/10.1103/PhysRevD.103.072007}{Phys. Rev. D \textbf{103}, 072007 (2021)}.

\bibitem{BESIII:2021uni}
M.~Ablikim \textit{et al.} [BESIII Collaboration],
Measurement of the $e^{+}e^{-}\rightarrow\omega\pi^{0}\pi^{0}$ cross section at center-of-mass energies from 2.0 to 3.08~GeV,
\href{https://doi.org/10.1103/PhysRevD.105.032005}{Phys. Rev. D \textbf{105}, 032005 (2022)}.

\bibitem{BESIII:2021aet}
M.~Ablikim \textit{et al.} [BESIII Collaboration],
Measurement of $e^+e^-\to\phi\pi^+\pi^-$ cross sections at center-of-mass energies from 2.00 to 3.08~GeV,
\href{https://doi.org/doi:10.1103/PhysRevD.108.032011}{Phys. Rev. D \textbf{108}, 032011 (2023)}.

\bibitem{BESIII:2022yzp}
M.~Ablikim \textit{et al.} [BESIII Collaboration],
Measurement of $e^{+}e^{-}\rightarrow\omega\pi^{+}\pi^{-}$ cross section at $\sqrt s = $ 2.000 to 3.080 GeV,
\href{https://doi.org/10.1007/JHEP01(2023)111}{JHEP \textbf{01}, 111 (2023)}; \href{https://doi.org/10.1007/JHEP03(2023)093}{\textbf{03}, 093 (E) (2023)}.

\bibitem{BESIII:2024qjv}
M.~Ablikim \textit{et al.} [BESIII Collaboration],
Measurement of $e^+e^-\to\omega\eta^\prime$ cross sections at $ \sqrt{s} $ = 2.000 to 3.080 GeV,
\href{https://doi.org/10.1007/JHEP07(2024)093}{JHEP \textbf{07}, 093 (2024)}.

\bibitem{ParticleDataGroup:2024cfk}
S.~Navas \textit{et al.} [Particle Data Group],
Review of particle physics,
\href{https://doi.org/10.1103/PhysRevD.110.030001}{Phys. Rev. D \textbf{110}, 030001 (2024)}.

\bibitem{Wang:2019mhs}
J.~Z.~Wang, D.~Y.~Chen, X.~Liu, and T.~Matsuki,
Constructing $J/\psi$ family with updated data of charmoniumlike $Y$ states,
\href{https://journals.aps.org/prd/abstract/10.1103/PhysRevD.99.114003}{Phys. Rev. D \textbf{99}, 114003 (2019)}.

\bibitem{Wang:2022jxj}
J.~Z.~Wang and X.~Liu,
Confirming the existence of a new higher charmonium \ensuremath{\psi}(4500) by the newly released data of $e^+e^-\to K^+K^-J/\psi$,
\href{https://journals.aps.org/prd/abstract/10.1103/PhysRevD.107.054016}{Phys. Rev. D \textbf{107}, 054016 (2023)}.
		
\bibitem{Wang:2023zxj}
J.~Z.~Wang and X.~Liu,
Identifying a characterized energy level structure of higher charmonium well matched to the peak structures in $e^+e^-\to \pi^+D^0D^{*-}$,
\href{https://www.sciencedirect.com/science/article/pii/S0370269324000157}{Phys. Lett. B \textbf{849}, 138456 (2024)}.

\bibitem{Peng:2024xui}
T.~C.~Peng, Z.~Y.~Bai, J.~Z.~Wang, and X.~Liu,
How higher charmonia shape the puzzling data of the $e^+e^-\to\eta J/\psi$ cross section,
\href{https://doi.org/10.1103/PhysRevD.109.094048}{Phys. Rev. D \textbf{109}, 094048 (2024)}.

\bibitem{Li:2021jjt}
Y.~S.~Li, Z.~Y.~Bai, Q.~Huang, and X.~Liu,
Hidden-bottom hadronic decays of $\Upsilon(10753)$ with a $\eta^{(\prime)}$ or $\omega$ emission, \href{https://doi.org/10.1103/PhysRevD.104.034036}{Phys. Rev. D \textbf{104}, 034036 (2021)}.

\bibitem{Bai:2022cfz}
Z.~Y.~Bai, Y.~S.~Li, Q.~Huang, X.~Liu, and T.~Matsuki,
$\Upsilon(10753)\rightarrow\Upsilon(nS)\pi^+\pi^- $decays induced by hadronic loop mechanism, \href{https://doi.org/10.1103/PhysRevD.105.074007}{Phys. Rev. D \textbf{105}, 074007 (2022)}.

\bibitem{Li:2022leg}
Y.~S.~Li, Z.~Y.~Bai, and X.~Liu,
Investigating the $\Upsilon(10753)\rightarrow\Upsilon (1^3D_J)\eta$ transitions, \href{https://doi.org/10.1103/PhysRevD.105.114041}{Phys. Rev. D \textbf{105}, 114041 (2022)}.

\bibitem{Liu:2024ets}
S.~D.~Liu, H.~D.~Cai, Z.~X.~Cai, H.~S.~Gao, G.~Li, F.~Wang, and J.~J.~Xie,
Production of $X_b$ via radiative transition of $\Upsilon{(10753)}$,
\href{https://doi.org/10.1103/PhysRevD.109.094045}{Phys. Rev. D \textbf{109}, 094045 (2024)}.

\bibitem{Liu:2023gtx}
S.~D.~Liu, Z.~X.~Cai, Z.~S.~Jia, G.~Li, and J.~J.~Xie,
Hidden-bottom hadronic transitions of $\Upsilon(10753)$,
\href{https://doi.org/10.1103/PhysRevD.109.014039}{Phys. Rev. D \textbf{109}, 014039 (2024)}.

\bibitem{Rosner:2001nm}
J.~L.~Rosner,
Charmless final states and $S$- and $D$- wave mixing in the $\psi^{\prime\prime}$,
\href{https://doi.org/10.1103/PhysRevD.64.094002}{Phys. Rev. D \textbf{64}, 094002 (2001)}.

\bibitem{Micu:1968mk}
L.~Micu,
Decay rates of meson resonances in a quark model,
\href{https://www.sciencedirect.com/science/article/abs/pii/055032136990039X}{Nucl. Phys. B\textbf{10}, 521 (1969)}.

\bibitem{LeYaouanc:1977gm}
A.~Le Yaouanc, L.~Oliver, O.~Pene, and J.~C.~Raynal,
Why is $\psi(4.414)$ so narrow?,
\href{https://www.sciencedirect.com/science/article/abs/pii/0370269377900624}{Phys. Lett. \textbf{72}B, 57 (1977)}.

\bibitem{Jacob:1959at}
M.~Jacob and G.~C.~Wick,
On the general theory of collisions for particles with spin,
\href{https://www.sciencedirect.com/science/article/abs/pii/000349165990051X}{Ann. Phys. (N.Y.) \textbf{7}, 404 (1959)}; \href{https://www.sciencedirect.com/science/article/pii/S0003491600960226}{\textbf{281}, 774 (2020)}.

\bibitem{Godfrey:1985xj}
S.~Godfrey and N.~Isgur,
Mesons in a Relativized Quark Model with Chromodynamics,
\href{https://doi.org/10.1103/PhysRevD.32.189}{Phys. Rev. D \textbf{32}, 189 (1985)}.

\bibitem{CMD-2:2000mlo}
R.~R.~Akhmetshin \textit{et al.} [CMD-2 Collaboration],
Study of the process $e^+e^-\to\pi^+\pi^-\pi^+\pi^-\pi^0$ with CMD-2 detector,
\href{https://doi.org/10.1016/S0370-2693(00)00937-0}{Phys. Lett. B \textbf{489}, 125 (2000)}.

\bibitem{CRYSTALBARREL:2001ldq}
A.~Abele \textit{et al.} [CRYSTAL BARREL Collaboration],
$4\pi$-decays of scalar and vector mesons,
\href{https://doi.org/10.1007/s100520100735}{Eur. Phys. J. C \textbf{21}, 261 (2001)}.

\bibitem{BaBar:2017dwm}
J.~P.~Lees \textit{et al.} [BaBar Collaboration],
Dalitz plot analyses of $J/\psi \to \pi^+ \pi^- \pi^0$, $J/\psi \to K^+ K^- \pi^0$, and $J/\psi \to K^0_S K^{\pm} \pi^{\mp}$ produced via $e^+ e^-$ annihilation with initial-state radiation,
\href{https://doi.org/10.1103/PhysRevD.95.072007}{Phys. Rev. D \textbf{95}, 072007 (2017)}.

\bibitem{CLEO:1999dln}
S.~Anderson \textit{et al.} [CLEO Collaboration],
Hadronic structure in the decay $\tau^-\to\pi^-\pi^0\nu_\tau$,
\href{https://doi.org/10.1103/PhysRevD.61.112002}{Phys. Rev. D \textbf{61}, 112002 (2000)}.

\bibitem{CrystalBarrel:1997mmp}
A.~Abele \textit{et al.} [Crystal Barrel Collaboration],
High-mass $\rho$-meson states from $\bar pd$-annihilation at rest into $\pi^-\pi^0\pi^0p_{\text{spectator}}$,
\href{https://doi.org/10.1016/S0370-2693(96)01552-3}{Phys. Lett. B \textbf{391}, 191 (1997)}.

\bibitem{DM2:1988xqd}
D.~Bisello \textit{et al.} [DM2 Collaboration],
The pion electromagnetic form factor in the time-like energy range $1.35\leq\sqrt s\leq2.4$ GeV,
\href{https://doi.org/10.1016/0370-2693(89)90060-9}{Phys. Lett. B \textbf{220}, 321 (1989)}.

\bibitem{CrystalBarrel:1999sjf}
A.~Abele \textit{et al.} [Crystal Barrel Collaboration],
Anti-proton proton annihilation at rest into $K^+K^-\pi^0$,
\href{https://doi.org/10.1016/S0370-2693(99)01191-0}{Phys. Lett. B \textbf{468}, 178 (1999)}.

\bibitem{Fukui:1988mp}
S.~Fukui \textit{et al.},
Vector resonances around 1.6 GeV of the $\eta \pi^+ \pi^-$ system in the $\pi^- p$ charge exchange reaction at 8.95 $\text{GeV}/c$,
\href{https://doi.org/10.1016/0370-2693(88)90500-X}{Phys. Lett. B \textbf{202}, 441 (1988)}.

\bibitem{Clegg:1993mt}
A.~B.~Clegg and A.~Donnachie,
Higher vector meson states produced in electron-positron annihilation,
\href{https://doi.org/10.1007/BF01555905}{Z. Phys. C \textbf{62}, 455 (1994)}.

\bibitem{SND:2014rfi}
V.~M.~Aulchenko \textit{et al.} [SND Collaboration],
Measurement of the $e^+e^- \to \eta\pi^+\pi^-$ cross section in the center-of-mass energy range 1.22--2.00 GeV with the SND detector at the VEPP-2000 collider,
\href{https://doi.org/10.1103/PhysRevD.91.052013}{Phys. Rev. D \textbf{91}, 052013 (2015)}.

\bibitem{Kurdadze:1983ys}
L.~M.~Kurdadze, M.~Y.~Lelchuk, E.~V.~Pakhtusova, V.~A.~Sidorov, A.~N.~Skrinsky, A.~G.~Chilingarov, Y.~M.~Shatunov, B.~A.~Shvarts, and S.~I.~Eidelman,
Measurement of the pion form factor at $640\leq\sqrt s\leq1400$ MeV,
\href{http://jetpletters.ru/ps/1499/article_22913.shtml}{JETP Lett. \textbf{37}, 733 (1983)};
\href{https://www.osti.gov/etdeweb/biblio/6584273}{Pisma Zh. Eksp. Teor. Fiz., \textbf{37}, 613}.
\end{thebibliography}
\end{document}